\documentclass{article}

\usepackage{arxiv}

\usepackage{hyperref}       
\usepackage{xcolor}
\hypersetup{colorlinks=true, linkcolor=blue, citecolor=blue}
\usepackage{amsfonts}       
\usepackage{amsmath}
\usepackage{amssymb}
\usepackage{nicefrac}       
\usepackage{microtype}      
\usepackage{lipsum}         
\usepackage{graphicx}
\usepackage{natbib}
\usepackage{doi}
\usepackage{siunitx}
\newcommand{\bm}[1]{{\mbox{\boldmath $#1$}}}

\title{Machine learning based dimension reduction for a stable modeling of periodic flow phenomena}

\newif\ifuniqueAffiliation
\uniqueAffiliationtrue

\ifuniqueAffiliation 
\author{Hiroshi Omichi\\
        Department of Mechanical Engineering,
        Keio University;\\
        Department of Mechanical and Aerospace Engineering,
        University of California, Los Angeles\\
        hiroshi.omichi@keio.jp\\
	\And
 Takeru Ishize \\
	Department of Mechanical Engineering,
	Keio University\\
	takeru.ishize@keio.jp \\
	\And
 Koji Fukagata \\
	Department of Mechanical Engineering,
	Keio University\\
	fukagata@mech.keio.ac.jp\\
}

\else
\usepackage{authblk}

\newbox{\orcid}\sbox{\orcid}{\includegraphics[scale=0.06]{orcid.pdf}} 
\author[1]{%
	\href{https://orcid.org/0000-0000-0000-0000}{\usebox{\orcid}\hspace{1mm}David S.~Hippocampus\thanks{\texttt{hippo@cs.cranberry-lemon.edu}}}%
}
\author[1,2]{%
	\href{https://orcid.org/0000-0000-0000-0000}{\usebox{\orcid}\hspace{1mm}Elias D.~Striatum\thanks{\texttt{stariate@ee.mount-sheikh.edu}}}%
}
\affil[1]{Department of Computer Science, Cranberry-Lemon University, Pittsburgh, PA 15213}
\affil[2]{Department of Electrical Engineering, Mount-Sheikh University, Santa Narimana, Levand}
\fi

\renewcommand{\headeright}{}
\renewcommand{\undertitle}{}

\begin{document}
\maketitle

\begin{abstract}
    In designing efficient feedback control laws for fluid flow,
    the modern control theory 
    can serve as a powerful tool
    if the model can be represented by
    a linear ordinary differential equation (ODE).
    However, it is generally 
    difficult
    to find such a linear model 
    for
    strongly nonlinear and high-dimensional fluid flow phenomena.
    In this study, we propose an autoencoder which maps the periodic flow phenomena into a latent dynamics governed by a linear ODE, referred to as a pseudo-symplectic Linear system Extracting AutoEncoder (LEAE).
    In addition to the normal functionality of autoencoder, 
    pseudo-symplectic LEAE emulates 
    a symplectic
    time integration scheme
    so that the Hamiltonian ({\it i.e.,} the pseudo-energy) of the latent variables is conserved.
    We demonstrate that the stability of the derived ODE is improved by considering the integration stepping forward and backward at the same time.
    Here, we consider the circular cylinder wake at $Re_D=100$ as a typical periodic flow phenomenon.
\end{abstract}


\section{Introduction}
Particularly when a target state of flow control locates far away from its base state, active flow control~\citep{park1994feedback,NF2014,JIS2020} is one of the promising techniques in terms of its efficacy and degrees of freedom in its design.
Among the active control strategies, 
model-free approaches~\citep{rabault2019artificial,WFJ2023} 
have strong merits in
their capability to follow the shifted dynamics and its transient process, although it is 
difficult
to interpret or generalize the constructed controllers~\citep{brunton2015closed}.
Model-based approaches 
have strong advantages also in that the existing control theories can be applied; when the model 
can be represented by
a linear ODE, there is a wide range of options like linear quadratic regulator even though they are only valid as long as the controlled states can be linearly approximated around the model.
However, as 
can 
easily 
be
imagined, it is difficult to find a proper model which governs fluid flow phenomena due to their high-dimensionality and strong nonlinearity.
In this context, dimension reduction with machine learning and modeling for the temporal evolution of low-dimensionalized variables ({\it i.e.,}latent variables) have been investigated~\citep{Milano2002,brenner2019perspective,hasegawa2020cnn,brunton2020special,farazmand2023tensor}; 
however,
with a 
na\"ive
dimension reduction,
the extracted latent dynamics 
will remain
highly nonlinear~\citep{FMZ2021}.

To address this problem, we propose an enhanced autoencoder named pseudo-symplectic Linear system Extracting AutoEncoder (LEAE) to compress flow field data into a latent 
space 
such that the latent dynamics is
governed by a linear ODE.
Inside the pseudo-symplectic LEAE, the dimension reduction 
is performed with an autoencoder, assuming
the time integration of latent variables 
using
the Crank-Nicolson scheme.
In particular, we propose
a method 
accounting for the symplectic property of the latent system by training
the temporal evolution 
in
forward and backward 
directions
simultaneously.
By so doing,
the Hamiltonian ({\it i.e.,} the pseudo-energy) of the latent variables 
$H=\frac{1}{2}\sum_i\varphi_i^2$
is conserved, where $\varphi_i$ denotes $i$-th component of the latent vector $\bm{\varphi}$.

In this study, we consider the iconic periodic flow phenomena in this field,
{\it i.e,} the periodic
 circular cylinder wake at the Reynolds number $Re_D=100$. 

\section{Methods}
\subsection{Dataset}
\label{subsec:DNS}

\begin{figure}[b]
    \centering
    \includegraphics[width=0.6\linewidth]{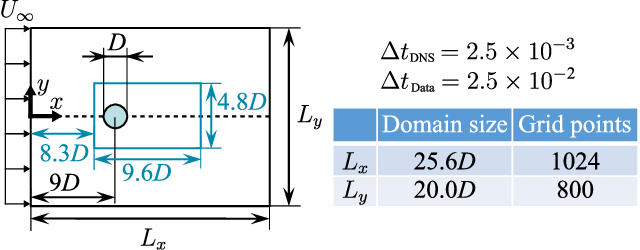}
    \caption{Computational configuration of DNS.}
    \label{fig:DNS}
\end{figure}

To prepare the training data, let us consider the two-dimensional flow around a circular cylinder at the Reynolds number based on its diameter ${\rm Re}_{D}=100$.
We use direct numerical simulation (DNS)~\citep{DNS_2D}, and the governing equations are the continuity and Navier-Stokes equations for incompressible flows,
\begin{align}
    \bm{\nabla} \cdot \bm u &= 0,\\
\frac{\partial \bm{u}}{\partial t}+\bm{\nabla} \cdot (\bm{uu})&=- \bm{\nabla} p+\frac{1}{{\rm{Re}}_D} \nabla^2 \bm{u},
\end{align}
where ${\bm u}=\{u,v\}$ and $p$ are the velocity vector and pressure, respectively.
Both quantities are nondimensionalized by the fluid density $\rho^{*}$, the diameter of cylinder $D^{*}$, and the uniform velocity $U_\infty^{*}$, where $(\cdot)^{*}$ 
denotes dimensional quantities.

The left side of Figure~\ref{fig:DNS} indicates the computational domain with 
the 
streamwise $(x)$ and transverse $(y)$ lengths of $25.6D$ and $20.0D$, respectively.
A uniform inflow velocity of $U_{\infty} = 1$, free-slip conditions on the upper and lower boundaries, and a convective boundary condition on the outflow boundary are applied,
while
the ghost-cell method~\citep{DNS_2D} 
is utilized
to impose the no-slip boundary condition on the surface.
A uniform 
Cartesian
grid 
system is used, and the number of grid points
in
each direction 
is
$\left( N_x, N_y\right) = \left(1024, 800\right)$, and the center of the cylinder locates at $x=9D$ downstream from the inflow boundary
 in $x$ direction 
and at the center in $y$ direction.
The time step is $\Delta t_{\rm DNS}=2.5\times10^{-3}$.

For training 
of
a machine learning model,
 the velocity fields $\left(u, v\right)$ inside the confined region depicted with the blue line in Figure~\ref{fig:DNS} 
 is used
 to extract 
 the
 key feature of 
 the
 cylinder wake.
The size of a snapshot is $\left(384 \times 192 \times 2\right)$, and we use the developed wake shedding of $2000$ snapshots with a time step of $\Delta t_{\rm Data}=2.5\times10^{-2}$ as the dataset for the machine learning model.

\subsection{Pesudo-symplectic linear system extraction autoncoder}
\subsubsection{Autoencoder}
\label{subsubsec:AE}

\begin{figure}[t]
    \centering
    \includegraphics[width=0.7\linewidth]{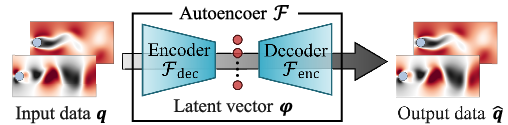}
    \caption{Schematic of a typical autoencoder.}
    \label{fig:AE}
\end{figure}

The machine learning model of this study is based on the consept of autoencoder~\citep{hinton2006reducing} to reduce the dimension of data.
In what follows, general description of autoencoders is detailed.
As illustrated in Figure~\ref{fig:AE}, an autoencoder $\cal{F}$ is comprised of an encoder $\cal{F}_{\rm enc}$ and a decoder $\cal{F}_{\rm dec}$ with a bottleneck in the middle.
As long as the autoencoder precisely replicates the input data ${\bm q} \in {\mathbb R}^{\mathit n_{\rm H}}$ and output it as $\hat{\bm q}\in {\mathbb R}^{\mathit n_{\rm H}}$, it can be said that the encoder maps the high dimensional data into
its low dimensional representation $\bm{\varphi} \in \mathbb{R}^{\mathit n_{\rm L}}$, {\it i.e.}, latent variables (vector), and the decoder 
has the
inverse function.
Note that $n_{\rm H}$ and $n_{\rm L}$ denote the dimension of the original and the low-dimensionalized data (usually, $n_{\rm H}\gg n_{\rm L}$), and
the latent variables $\bm{\varphi}$ probably hold essential characteristics of the original dynamics of ${\bm q}$.
This conversion can be formulated as
\begin{align}
    {\bm q} \approx \hat{\bm q} = \cal{F}(\bm q; \bm w) = \cal{F}_{\rm dec}(\bm{\varphi}) = \cal{F}_{\rm dec}(\cal{F}_{\rm enc}(\bm{\bm q})),
    \label{eq:AE}
\end{align}
where $\bm w$ 
denotes the
weights
inside the machine learning model 
 to be optimized so that
\begin{align}
    {\bm w}={\rm argmin}_{{\bm w}}{\parallel \bm{q}-\hat{\bm q} \parallel}_2.
\end{align}

In terms of the network architecture, we follow~\cite{hasegawa2020cnn} and employ nonlinear neural networks, {\it i.e.}, Convolutional Neural Network~\citep{LBBH1998} (CNN) and Multi-layer perception~\citep{MLP}.

\subsubsection{Linear ODE layer}
\label{subsubsec:LODE}
After the dimension reduction, 
the
temporal evolution of the latent variables have to be modeled to construct a reduced order model.
In this study, the temporal evolution is predicted with 
what we call
the linear ODE (LODE) layer,
 which follows a temporal discretization of linear ODEs
 using the Crank-Nicolson scheme.
Considering a linear ODE,
\begin{align}
    \dot{\bm{\varphi}}={\bm A}{\bm{\varphi}},
    \label{eq:ODE}
\end{align}
where $\bm{\varphi} \in \mathbb{R}^{\mathit n_{\rm L}}$ corresponds to the latent vector in section \ref{subsubsec:AE}, and $\bm A$ is the coefficient matrix to be optimized through the training process, equation~(\ref{eq:ODE}) can be discretized with Crank-Nicolson scheme as
\begin{align}
    & \frac{\bm{\varphi}(t+\Delta t)-\bm{\varphi}(t)}{\Delta t} = \frac{A[\bm{\varphi}(t+\Delta t)+\bm{\varphi}(t)]}{2}\\
    & \Rightarrow \bm{\varphi}(t+\Delta t) = (2I+\Delta tA)(2I-\Delta tA)^{-1}\bm{\varphi}(t).
    \label{eq:CN_ODE}
\end{align}
Regarding $\bm{\varphi}(t)$ and $\bm{\varphi}(t+\Delta t)$ as input and output, respectively, we can use a single MLP layer corresponding to $n_{\rm L} \times n_{\rm L}$ matrix $\bm A$.

\subsubsection{Overview of pseudo-symplectic linear system extraction autoencoder}
\label{subsubsec:LEAE}

\begin{figure}[t]
    \centering
    \includegraphics[width=0.7\linewidth]{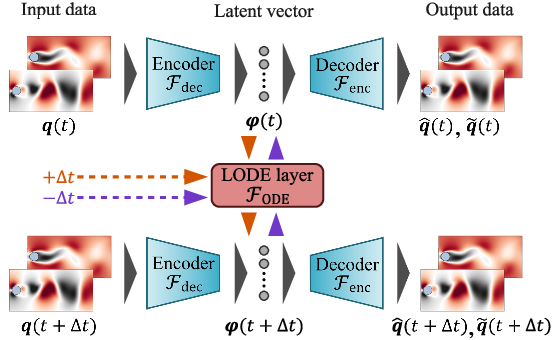}
    \caption{Schematic of pseudo-symplectic linear system extraction autoencoder.}
    \label{fig:LEAE}
\end{figure}

Here, the overview of our machine learning model, pseudo-symplectic linear system extraction autoencoder (LEAE), is summarized in Figure~\ref{fig:LEAE}.
The pseudo-symplectic LEAE basically consists of the encoder, the decoder, and the LODE layer in the middle.
Both of the encoders completely share the weights, and both of the decoders do as well, which means that only a single encoder and decoder are trained here.
Needless to say, the encoder and the decoder are trained to duplicate the original snapshot as precisely as possible, but what we want to highlight here is the usage of snapshots mainly to train LODE layer.
LODE layer is trained to not only predict ${\bm \varphi}(t+\Delta t)$ from ${\bm \varphi}(t)$ (the orange arrows) but also ${\bm \varphi}(t)$ from ${\bm \varphi}(t+\Delta t)$ (the purple arrows).
By doing so, LODE layer 
can be trained so as to satisfy the time reversal symmetry, which 
is the important property of symplectic integrators that must be satisfied in the time integration of Hamiltonian system \citep{Aceto}.
In sum,
the present training process consists of 
four types of operations:
\begin{align}
    \begin{cases}
    \hat{\bm q}(t) &= \mathcal{F}_{\rm dec}(\mathcal{F}_{\rm enc}({\bm q}(t))) = \mathcal{F}_{\rm dec}({\bm \varphi}(t))\\
    \hat{\bm q}(t+\Delta t) &= \mathcal{F}_{\rm dec}(\mathcal{F}_{\rm enc}({\bm q}(t+\Delta t))) = \mathcal{F}_{\rm dec}({\bm \varphi}(t+\Delta t))\\
    \Tilde{\bm q}(t+\Delta t) &= \mathcal{F}_{\rm dec}(\mathcal{F}_{\rm ODE}({\bm \varphi}(t),\Delta t))\\
    \Tilde{\bm q}(t) &= \mathcal{F}_{\rm dec}(\mathcal{F}_{\rm ODE}({\bm \varphi}(t+\Delta t),-\Delta t))\\
    \end{cases}
    ,
    \label{eq:Symp}
\end{align}
where $\Delta t$ is a time step.
While the former two equations contribute to guaranteeing that $\cal{F}_{\rm enc}$ and $\cal{F}_{\rm dec}$ are providing the nonlinear mappings at exactly the same time indices, the latter two commit to learning the temporal evolution stepping forward and backward, and all operations are taken simultaneously.
Therefore, the loss function is 
defined as
\begin{align}
    {\cal L} = {\parallel \bm{q}(t)-\hat{\bm q}(t) \parallel}_2 + {\parallel \bm{q}(t+\Delta t)-\hat{\bm q}(t+\Delta t) \parallel}_2 + {\parallel \bm{q}(t+\Delta t)-\Tilde{\bm q}(t+\Delta t) \parallel}_2 + {\parallel \bm{q}(t)-\Tilde{\bm q}(t) \parallel}_2.
    \label{eq:Loss}
\end{align}

\section{Results}
\label{sec:Result}

\begin{figure}[t]
    \centering
    \includegraphics[width=0.9\linewidth]{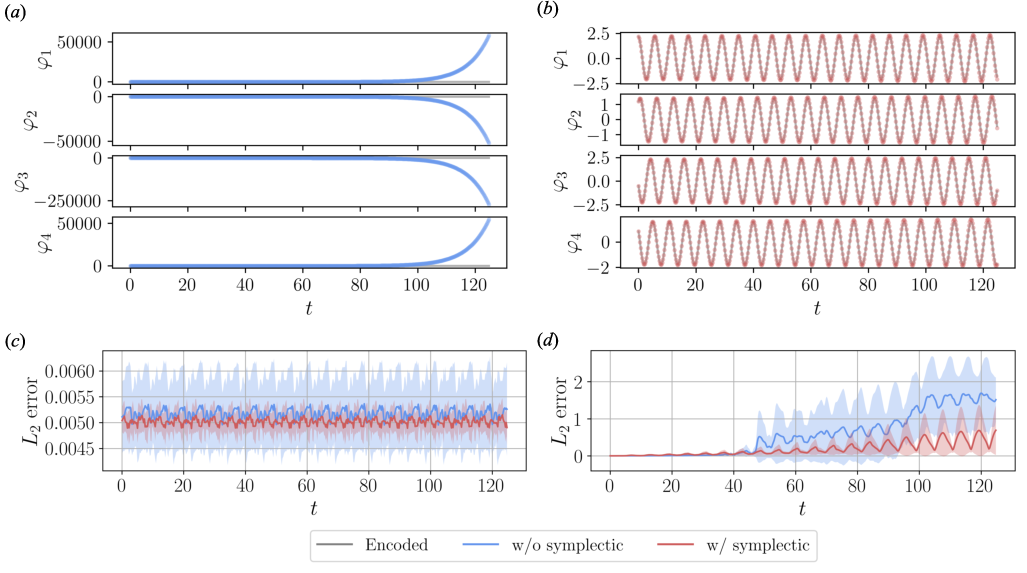}
    \caption{The performances of the naive LEAE (blue plots) and the pseudo-symplectic LEAE (red plots). $(a)$, $(b)$ reproduced latent trajectories, $(c)$ $L_2$ error of reconstructed fields, and $(d)$ $L_2$ error of decoded fields from the reproduced latent trajectories.}
    \label{fig:Res}
\end{figure}

In this section, the efficacy of the pseudo-symplectic manner is summarized 
by
comparing the pseudo-symplectic LEAE and a na\"ive LEAE.
Here, the na\"ive LEAE is a form of LEAE which excludes the temporally inverse procedure denoted as the purple arrows in figure~\ref{fig:LEAE}.

Figure~\ref{fig:Res} visualizes the performances of the na\"ive LEAE (blue plots) and the pseudo-symplectic LEAE (red plots).
In figure~\ref{fig:Res} $(a)$ and $(b)$, the latent variables encoded from flow fields (gray lines) and ones predicted with numerical integration of the derived ODEs (colored lines) are compared.
While the predicted trajectories blow up in the case of the na\"ive LEAE, the encoded flow fields and the result of numerical integration show a nice agreement in the case of the pseudo-symplectic 
LEAE.
Note that the initial values for the cases of the numerical integration are given through encoding the initial flow fields.
In figure~\ref{fig:Res} $(c)$ and $(d)$, both types of LEAEs are evaluated in terms of $L_2$ error of decoded fields at each snapshot, and means and standard deviations taken over three-fold cross validation are presented.
Shown in figure~\ref{fig:Res} $(c)$ is the errors of the reconstructed flow fields through the autoencoder.
Please note 
that
no temporal evolution is considered here; the reconstruction procedure corresponds to the horizontal flow in figure~\ref{fig:LEAE}, and the error is exactly the same as the first term of the equation~(\ref{eq:Loss}).
The errors including those along
the temporal evolution 
are
presented
in figure~\ref{fig:Res} $(d)$.
Here, decoded fields from the predicted trajectories in figure~\ref{fig:Res} $(a)$ and $(b)$ are assessed.
The above two evaluations demonstrate that the pseudo-symplectic LEAE is a
substantially
 improved method compared to the na\"ive one.

\section{Conclusions}
In this study, we proposed a pseudo-symplectic Linear system Extraction AutoEncoder (LEAE) to exract a low-dimensionalized dynamics which is governed by a linear ODE. 
It has been shown that the pseudo-symplectic LEAE has the ability to derive an ODE which models a periodic latent dynamics more precisely than a na\"ive LEAE.
However, in the present study, we only considered the case
where
the ODE strongly adheres to a limit-cycle corresponding to the periodic wake shedding, and this modeling is imaginably not applicable for cases where flow states are shifting.
Toward an effective flow control framework, there is room for further research into a modeling capable of following transient processes due to control inputs.

\section*{Acknowledgments}
This work was supported by JSPS KAKENHI Grant Number 21H05007.
The authors acknowledge Mr. Shoei Kanehira (Keio University) for fruitful discussion and comments.


\end{document}